\documentclass[twocolumn]{aastex6}		
\pdfoutput=1
\usepackage{natbib}

\shorttitle{New age and physical properties of HD 203030B}
\shortauthors{Miles-P\'aez et al.}

\begin{document}

\title{The Prototypical Young L/T-Transition Dwarf HD 203030B Likely Has Planetary Mass}

\author{Paulo A. Miles-P\'aez\altaffilmark{1,2}, Stanimir Metchev\altaffilmark{1,3,4}, Kevin L. Luhman\altaffilmark{5,6}, Massimo Marengo\altaffilmark{7} \& Alan Hulsebus\altaffilmark{7}}
\altaffiltext{1}{Department of Physics \& Astronomy and Centre for Planetary Science and Exploration, The University of Western Ontario, London, Ontario N6A 3K7, Canada  (\texttt{ppaez@uwo.ca})}
\altaffiltext{2}{Steward Observatory and Department of Astronomy, University of Arizona, 933 N. Cherry Avenue, Tucson, AZ 85721, USA}
\altaffiltext{3}{Department of Physics \& Astronomy, Stony Brook University, Stony Brook, NY 11794--3800, USA}
\altaffiltext{4}{Visiting Astronomer at the NASA Infrared Telescope Facility}
\altaffiltext{5}{Department of Astronomy and Astrophysics, The Pennsylvania State University, University Park, PA 16802, USA}
\altaffiltext{6}{Center for Exoplanets and Habitable Worlds, The Pennsylvania State University, University Park, PA 16802, USA}
\altaffiltext{7}{Department of Physics and Astronomy, Iowa State University, Ames, IA 50011, USA}

\begin{abstract}
Upon its discovery in 2006, the young L7.5 companion to the solar analog HD 203030 was found to be unusual in being $\approx$200 K cooler than older late-L dwarfs.  HD~203030B offered the first clear indication that the effective temperature at the L-to-T spectral type transition depends on surface gravity: now a well-known characteristic of low-gravity ultra-cool dwarfs.  An initial age analysis of the G8V primary star indicated that the system was 130--400~Myr old, and so the companion between 12--31~$M_{\rm Jup}$.  Using moderate resolution near-infrared spectra of HD~203030B, we now find features of very low gravity comparable to those of 10--150 Myr-old L7--L8 dwarfs. We also obtained more accurate near infrared and {\sl Spitzer}/IRAC photometry, and find a $(J-K)_{\rm MKO}$ color of $2.56\pm0.13$ mag---comparable to those observed in other young planetary-mass objects---and a luminosity of log($L_{\rm bol}/L_{\odot}$)$\,=\,-4.75\pm0.04$ dex. We further reassess the evidence for the young age of the host star, HD 203030, with a more comprehensive analysis of the photometry and updated stellar activity measurements and age calibrations.  Summarizing the age diagnostics for both components of the binary, we adopt an age of 100 Myr for HD~203030B and an age range of 30--150 Myr. Using cloudy evolutionary models, the new companion age range and luminosity result in a mass of 11~$M_{\rm Jup}$ with a range of 8--15~$M_{\rm Jup}$, and an effective temperature of $1040\pm50$ K. 
\end{abstract}


\section{INTRODUCTION}

Substellar objects cool and dim continuously throughout their lifetimes.  Lacking the luminosity---effective temperature ($T_{\rm eff}$) relation of main sequence stars, substellar objects present formidable challenges to resolving the degeneracies among their fundamental parameters with age.  Brown dwarfs with known ages, masses, or metallicities---often referred to as ``benchmark'' substellar objects \citep{2006MNRAS.368.1281P,2008ApJ...689..436L}---are thus a valuable resource in studying substellar properties.  

The interest in cool substellar benchmarks is elevated because of their similarities to non-irradiated extrasolar giant planets. The set of directly imaged planetary-mass companions, all of which have ages younger than 500~Myr, show similar characteristics of young L or T dwarfs: very red optical to mid-infrared (mid-IR) colors indicative of dust-rich atmospheres, weak alkali absorption lines indicative of low surface gravities, and a reduced methane-to-carbon monoxide ratio suggesting enhanced vertical mixing \citep{2006ApJ...651.1166M, 2009ApJ...694L.148L, 2011ApJ...732..107S, 2011ApJ...743..191M, 2012ARA&A..50...65L,2013AJ....145....2F,2013ApJ...777L..20L}.

The list of planetary-mass objects ($\lesssim13\,M_{\rm Jup}$, for solar metallicities) and young brown dwarfs reported during the last decade is long, with some excellent candidates to investigate the likely evolution of the L/T transition with age. Some of these objects include the young companions HD 203030B \citep[L7.5, 130--400 Myr;][]{2006ApJ...651.1166M}, HN PegB \citep[T2.5, 100--500~Myr;][]{2007ApJ...654..570L}, LP 261--75B \citep[L6, 100--200 Myr;][]{2006PASP..118..671R}, GU Psc b \citep[T3.5, 70--130 Myr;][]{2014ApJ...787....5N}, or VHS J1256--12 \citep[L7, 150--300 Myr;][]{2015ApJ...804...96G}, or the very low-mass brown dwarfs PSO J318.5--22 \citep[L7, $23\pm3$ Myr;][]{2013ApJ...777L..20L,2016ApJ...819..133A}, 2MASS J1119--11AB \citep[L7, 10$\pm$3 Myr;][]{2016ApJ...821L..15K,2017ApJ...843L...4B} or WISEA J1147--20 \citep[L7, 10$\pm$3 Myr;][]{2016ApJ...822L...1S}. 

HD~203030B was discovered as a co-moving companion to its host star by \citet{2006ApJ...651.1166M}. It was the first to be recognized to show a characteristically cooler $T_{\rm eff}$ than older field brown dwarfs at the same spectral type: subsequently confirmed and established as a characteristic of low-gravity ultra-cool dwarfs \citep[e.g.,][]{2007ApJ...654..570L,2011ApJ...735L..39B,2013ApJ...777L..20L,2015ApJ...804...96G}. The discovery paper included a $J$ band magnitude with an uncertainty of $\pm$0.55 mag and lacked $J$ or $H$ band spectra, so precluding a better physical understanding of HD 203030B.

In this paper we report improved near-infrared photometry, new low- and moderate-resolution 0.85--2.5$\mu$m spectra, and 3.6--8.0$\mu$m {\sl Spitzer} photometry to accurately determine the physical properties of HD 203030B. These data---in combination with the gravity classification framework for ultra-cool dwarfs developed in the last years \citep{2013ApJ...772...79A,2013MNRAS.435.2650C,2014A&A...562A.127B}---allowed us to update the likely age, mass, and $T_{\rm eff}$ of HD 203030B. We describe our new data and their reduction in Section \ref{obs}. The revised age and physical properties of HD 203030B are discussed in Section \ref{results}, and our main conclusions are summarized in Section \ref{conclusions}.

\begin{deluxetable*}{lccccccccc}
\tablewidth{0pt}
\tabletypesize{\footnotesize}
\tablecaption{Summary of Observations of HD 203030B \label{tab_observations}}
\tablehead{
	\multicolumn{2}{l}{Object} &\multicolumn{2}{c}{UT Date} &\multicolumn{2}{c}{Band/Wavelength ($\mu$m)} & 
	\multicolumn{2}{c}{Instrument/Telescope} & \multicolumn{2}{r}{Reference}}
\startdata
\multicolumn{10}{c}{\textbf{Photometry}} \\
\multicolumn{2}{l}{HD 203030B} & \multicolumn{2}{c}{2011 Jul 17} & \multicolumn{2}{c}{MKO $J, H$} & \multicolumn{2}{c}{WIRC/Palomar} & \multicolumn{2}{r}{1} \\
 \multicolumn{2}{l}{ }& \multicolumn{2}{c}{2004 Jun 26} & \multicolumn{2}{c}{CIT $K_s$} & \multicolumn{2}{c}{PHARO/Palomar} & \multicolumn{2}{r}{2} \\
 \multicolumn{2}{l}{}& \multicolumn{2}{c}{2007 Nov 15} & \multicolumn{2}{c}{channels 1--4} & \multicolumn{2}{c}{IRAC/{\it Spitzer}} & \multicolumn{2}{r}{1} \\
\\
\multicolumn{10}{c}{\textbf{Spectroscopy}}\\
\hline
\multicolumn{1}{l}{Object} &Instrument & Date & Filter & Slit Width & Wavelength &$R=\lambda/\Delta\lambda$ & Exposure & SNR & \multicolumn{1}{r}{Reference}\\
\multicolumn{1}{l}{}           &                  &          &          &($\arcsec$)&  ($\mu$m)   &                                              & (min)          &        &\multicolumn{1}{r}{} \\
\hline
\multicolumn{1}{l}{HD 203030B} & IRTF/SpeX       & 2010 Jul 21 & \nodata & 0.80  & 0.85--2.5& 120  & 96  & \multicolumn{1}{r}{12} & 1 \\
\multicolumn{1}{l}{}	  & { Keck/NIRC2}    &{ 2005 Jul 14} & -- & { 0.08} & { 2.03--2.36} &{ 1300} & { 10}  & \multicolumn{1}{r}{{ 25}}  & { 2}\\
\multicolumn{1}{l}{}	  & Keck/NIRSPEC    & 2008 Jul 8 & N3 & 0.38 &1.15--1.38 & 2300 & 110 & \multicolumn{1}{r}{19} & 1 \\
\multicolumn{1}{l}{}	  & Keck/NIRSPEC    & 2008 Jul 8 & N5 & 0.76 & 1.50--1.78 &1500 & 40  & \multicolumn{1}{r}{12}  & 1\\
\enddata
\tablerefs{1. This paper; 2. \citet{2006ApJ...651.1166M}.}
\end{deluxetable*}
\section{OBSERVATIONS AND DATA REDUCTION}\label{obs}

The discovery paper of HD 203030B included near-IR photometry and an $R\approx1300$ $K$-band spectrum obtained with NIRC2 on Keck II. To the existing data we add more comprehensive and/or higher-resolution near-IR spectroscopy, and near-IR and {\it Spitzer}/IRAC photometry.  All photometric and spectroscopic observations are summarized in Table~\ref{tab_observations}.

\begin{figure*}
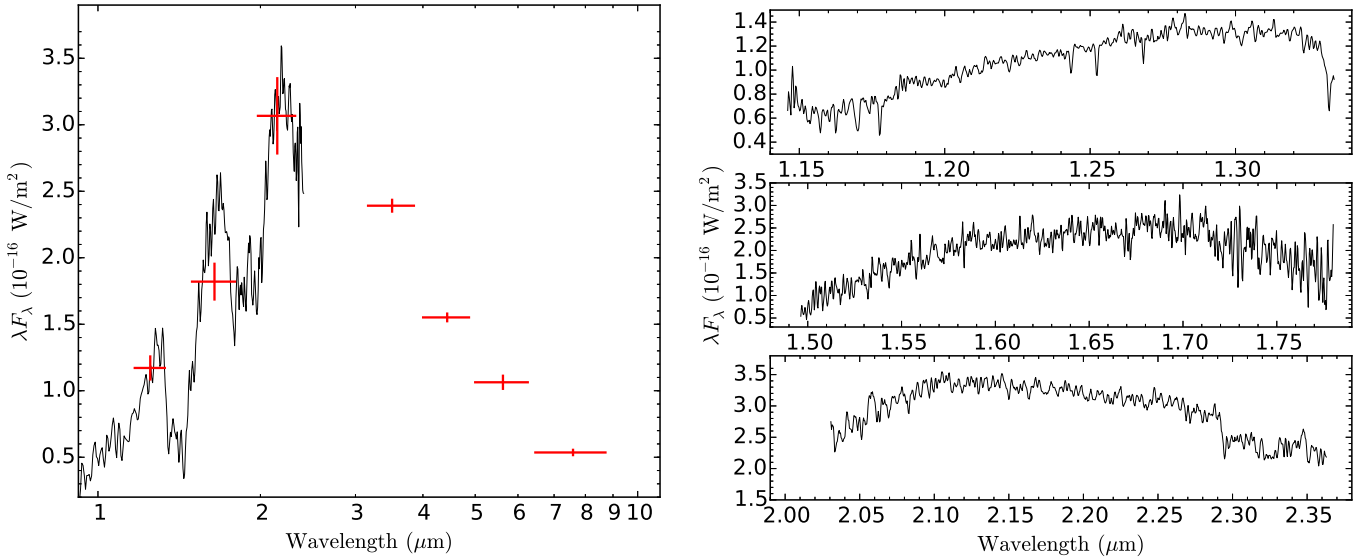

\centering
\gridline{\leftfig{fig_SED1.eps}{0.49\textwidth}{} \rightfig{fig_SED2.eps}{0.49\textwidth}{}}
\caption{{\sl Left:} IRTF/SpeX prism spectra of HD 203030B (black; Section \ref{sec_hd203030b_spex}), along with Palomar/WIRC (Section \ref{sec_palomar_photometry}) and {\it Spitzer}/IRAC (Section \ref{irac}) photometry (red data points with error bars). Horizontal bars indicate the width of the passband of each filter. {\sl Right:} $J$- (top) and $H$-band (middle) moderate resolution spectra taken with Keck/NIRSPEC (Section \ref{kecknir}). We have also included the $K$-band spectrum (bottom) taken from \citet{2006ApJ...651.1166M}. 
\label{figseds}}
\end{figure*}

\subsection{New Near-IR Photometry and Spectroscopy}

\subsubsection{WIRC/Hale Photometry of HD 203030B
\label{sec_palomar_photometry}}

The near-IR photometry of HD~203030B presented in \citet{2006ApJ...651.1166M} was obtained from adaptive optics observations on the Hale and Keck telescopes, with the primary used as the adaptive optics guide star.  Given the 11.9$\arcsec$ separation of HD~203030B from the primary, the shorter wavelength ($J$ and $H$)  photometry was strongly affected by anisoplanatic distortions of the PSF.  Notably, the Keck $J$-band photometry had an error of $\pm$0.55~mag.

We obtained more reliable seeing-limited MKO $J$ and $H$ photometry of HD~203030B with the WIRC camera \citep{2003SPIE.4841..451W} on the Hale telescope in 2011 July. WIRC has a 2048-square Rockwell Hawaii-II near-IR detector with a plate scale of 0.248 \arcsec/pixel, which yields a field of view of $\sim8.7\arcmin\times8.7\arcmin$. We collected 11 images in the $J$ band and 8 in the $H$ band---following a dither pattern---with individual exposure times of 60 s. The average seeing during these  observations was $\sim$1\arcsec. 

We used the Image Reduction and Analysis Facility software (IRAF) for sky subtraction, flat fielding, alignment, and median combination of our data following standard routines. Despite the wide angular separation of the companion, there was a small sky gradient near HD 203030B due to the bright halo of the primary star. We subtracted that gradient by fitting a 2D profile to the halo of the primary with the IRAF task {\sc ellipse}, and subtracted the best model to the images. We checked that the sky level and its standard deviation in the regions close to HD 203030B were similar to those of regions not affected by the halo of the primary. For each filter, we performed aperture circular photometry of HD 203030B using a radius of $1\farcs2$ and three sky annulus with inner radii of $1\farcs3$, $1\farcs8$ and $2\farcs2$ and a width of $0\farcs5$. We computed the average value of these three measurements and their standard deviation as the final value and uncertainty of the instrumental magnitude of HD 203030B. We calibrated the photometric measurements with respect to those of 27 ($J$) and 23 ($H$) nearby stars with magnitudes 0.2--2.0 mag brighter than those of HD 203030B and with precise  photometry ($\pm$0.02--0.06 mag) from 2MASS \citep{2006AJ....131.1163S}. We converted the 2MASS magnitudes of these calibration stars to the MKO system \citep{2002PASP..114..169S} following the relations given in \citet{2009MNRAS.394..675H}, and fit first-order polynomials between the instrumental and MKO magnitudes of the calibrators. Finally we used these fits to calibrate the magnitude of HD 203030B. Our magnitude determinations for the $J$ and $H$ bands (MKO) are $18.77\pm0.08$ mag and $17.57\pm0.08$ mag, respectively. The uncertainties were derived by adding in quadrature the uncertainty of the instrumental magnitude and the standard deviation of the residuals of the linear fit for each filter.

We also collected $K$-band photometry with WIRC. However, the counts in the stellar halo were well into the non-linear response regime for the detector, and did not allow for an accurate photometric calibration. We adopt the 2MASS-like $K_s$ filter measurement from \citet[][$16.21\pm0.10$ mag]{2006ApJ...651.1166M}. From the relations presented in \citet{2012ApJS..201...19D}, we find $0.00 \leq K_{\rm MKO}-K_{\rm 2MASS}\leq 0.09$~mag for L7-L8 dwarfs. So, we do not apply any correction to the $K_{\rm 2MASS}$ magnitude of HD 203030B and use it as a representative value of the $K_{\rm MKO}$ magnitude.

\subsubsection{IRTF/SpeX Spectra of HD 203030B
\label{sec_hd203030b_spex}}

We observed HD 203030B with SpeX \citep{2003PASP..115..362R} in prism mode and the $0\farcs8\times15\farcs0$ slit for 2.6 hours on 21 July 2010.  Fifty-two individual 180~s exposures were recorded following a standard A--B--B--A nodding sequence along the slit.  The A0 star HD~212734 was observed for telluric correction. Flat-field and argon lamp exposures were recorded immediately after each set of target and standard star observations.

All reduction steps, flat-fielding, bad pixel correction, wavelength calibration, and extraction were carried out with the SpeXtool package version 3.2 \citep{2003PASP..115..389V,2004PASP..116..362C}. The reduction of the HD~203030B spectra was challenging because of the faintness of the target and the proximity to the primary HD~203030A. The primary star produced a bright background and a visible trace in the raw data. During the reduction we disregarded 20 of the 52 exposures, since the spectral trace of HD~203030B was not visible. To improve the spectral extraction in the remaining 32 exposures, we averaged them in groups of eight taken at the same nodding position.  The extraction aperture had a width of 1$\farcs$2, which is 1.4--1.6 times the full-width-at-half-maximum (FWHM) of the spatial profile of the spectrum, with the median pixel value of the neighboring regions (where not contaminated by the host star) used as an estimate of the local background.  Given the low signal-to-noise ratio (SNR) of the data we did not use optimal extraction, but simply summed the flux across the aperture (i.e., using uniform weights).  The reduced spectrum was flux-calibrated using the photometry presented in Section~\ref{sec_palomar_photometry}, and smoothed to the instrument resolution using a Savitzky-Golay smoothing algorithm. The final SpeX spectrum of HD~203030B has a spectral resolution ($R$) of $\sim120$ and covers the range  0.85--2.5~$\micron$.

\subsubsection{Keck/NIRSPEC Spectra of HD 203030B\label{kecknir}}

We obtained moderate-resolution ($R\approx2200$) spectra of HD~203030B using NIRSPEC \citep{1998SPIE.3354..566M} on the Keck II telescope on 8 July 2008. Spectra of HD~203030B were taken through the N3 (1.143--1.375~$\micron$) and N5 (1.413--1.808~$\micron$) filters with the two- ($0\farcs38$) and four-pixel ($0\farcs76$) wide slits, respectively. Standard stars and arc lamps were observed after each science target.

We performed preliminary data reduction by using the REDSPEC pipeline \citep{2003ApJ...596..561M}. Individual exposures were flat-fielded, rectified, and wavelength calibrated. Optimal extraction of the spectra was done with the IRAF {\sc apall} package. After correcting for telluric absorption, the individual spectra were median combined. The resultant spectra were flux calibrated using the SpeX spectra presented herein (Section~\ref{sec_hd203030b_spex}).  

The final Spex and NIRSPEC spectra of HD~203030B from Section \ref{sec_hd203030b_spex} and this Section are shown in Figure~\ref{figseds}.  The displayed NIRSPEC $H$ band spectrum of HD~203030B has been Savitzky-Golay smoothed to approximately match the resolution of the 4~pixel wide spectroscopic slit.

\begin{figure}
\centering
\includegraphics[width=0.47\textwidth]{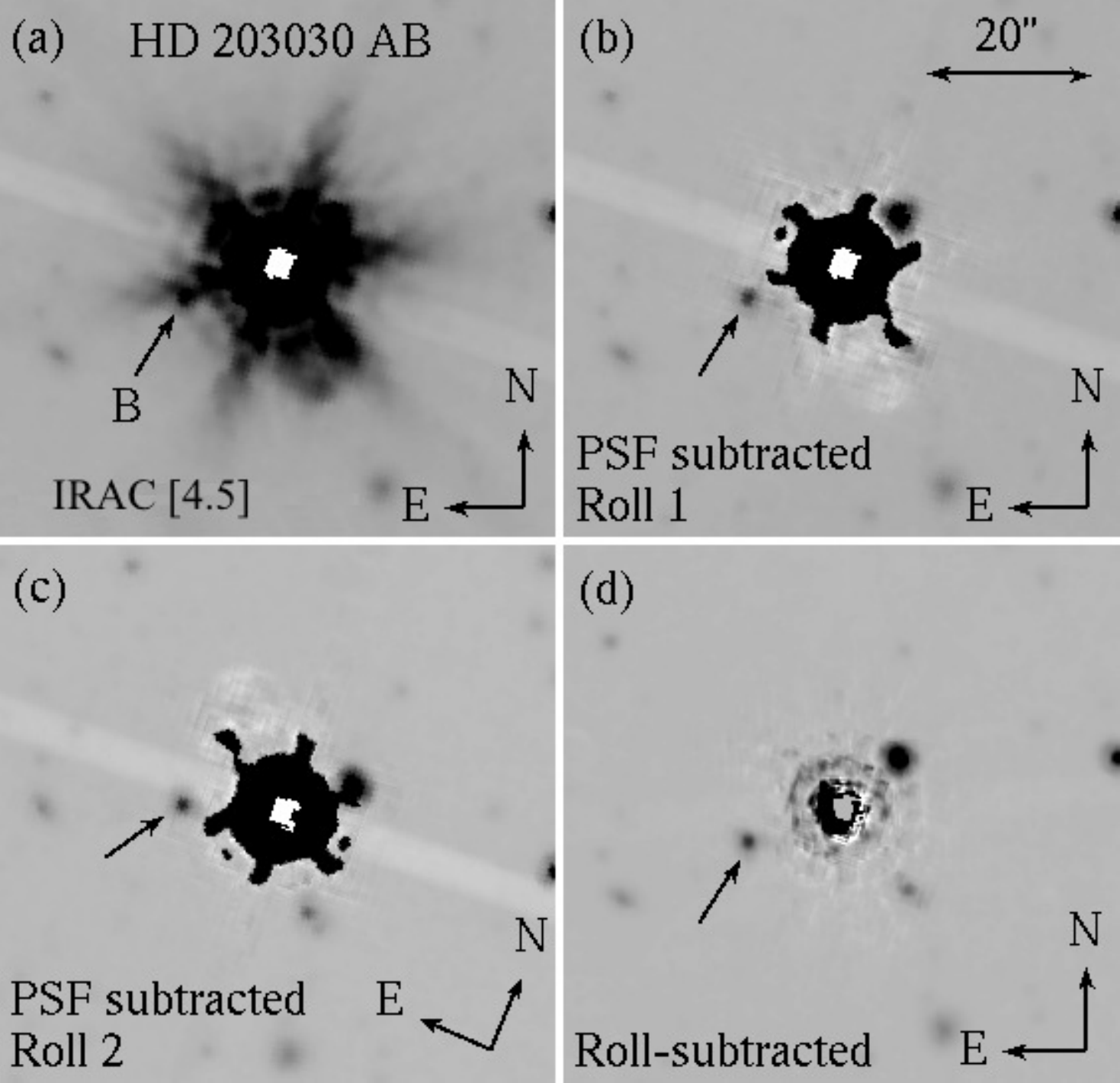}
\caption{{\it Spitzer}/IRAC channel 2 images of the HD~203030A/B pair.  Panel (a) shows the 480~s of median-combined exposures from the first imaging epoch.  Panels (b) and (c) show the PSF-subtracted median-combined exposures from the two epochs with telescope roll angle differing by 23$\degr$.  Panel (d) shows the roll-angle subtracted, combined 960~s image.
\label{fighd203030irac}}
\end{figure}

\subsection{IRAC Photometry of HD 203030B}\label{irac}

We used {\it Spitzer} to collect IRAC photometry of HD 203030B under program GO 40489. We did not attempt to obtain IRS spectroscopy since it is prohibitively challenging due to the separation (11$\farcs$9) between HD 203030B and its host star.  

HD~203030B is $\sim$8~mag fainter than its host star in the IRAC bands, at an angular separation of only six diffraction beam widths in the IRAC 8~$\micron$ channel.  To enable removal of contamination from the host star's halo, we observed the HD~203030A/B pair at two distinct epochs, with telescope roll angles differing by 23$\degr$.  The data were taken in high dynamic range (HDR) mode in sequences of 1.2 and 30~s exposures at each of 16 dithered pointings along a medium-scale ($\sim$120$\arcsec$ wide) half-subpixel spiral pattern.  The HDR mode enabled the recording of unsaturated images of the host star, which were subsequently used for image registration.  The total  integration time in 30~s exposures was 960~s per channel.

The data were reduced with version S18.7.0 of the IRAC Pipeline, which produced Basic Calibrated Data (BCD) frames and data quality masks for each individual full frame and subarray exposure.  The roll-angle subtraction of the primary PSF was done following the procedure described in \citet{2009ApJ...700.1647M}, using the IRACProc package \citep{2006SPIE.6270E..20S}. Figure~\ref{fighd203030irac} depicts the PSF and roll-angle subtraction steps for the IRAC channel 2 images.

The flux of HD~203030B was measured from the final roll-subtracted, combined images, as in panel (d) of Figure~\ref{fighd203030irac}.  In each channel we summed the flux in an $r=2$~pix (2$\farcs$4) aperture centered on HD~203030B, and compared that to the fluxes of three isolated, bright field stars within identical apertures.  Aperture corrections were estimated as the differences between the fluxes of the field stars in $r=10$~pix and $r=2$~pix apertures.  The flux errors and the aperture correction errors were propagated to obtain the photometric errors for HD~203030B.  The (Vega) photometry for the four IRAC channels and the near-IR photometry derived in Section \ref{sec_palomar_photometry} are listed in Table~\ref{tab_spitzer_phot}, and are plotted in Figure \ref{figseds} (left). 

\begin{deluxetable}{lccccr}
\tabletypesize{\footnotesize}
\tablewidth{0pt}
\tablecaption{Measurements and determined physical parameters of HD 203030B
\label{tab_spitzer_phot}}
\tablehead{\multicolumn{2}{l}{Property} & \multicolumn{2}{c}{Measurement} & \multicolumn{2}{r}{Reference}}
\startdata 
\multicolumn{6}{c}{Photometry} \\
\hline \hline
\multicolumn{2}{l}{$J_{}$} & \multicolumn{2}{c}{$18.77\pm0.08$ mag} & \multicolumn{2}{r}{1}\\
\multicolumn{2}{l}{$H$} & \multicolumn{2}{c}{$17.57\pm0.08$ mag} & \multicolumn{2}{r}{1}\\
\multicolumn{2}{l}{$K_s$} & \multicolumn{2}{c}{$16.21\pm0.10$ mag} & \multicolumn{2}{r}{2}\\
\multicolumn{2}{l}{[3.6]} & \multicolumn{2}{c}{$14.99\pm0.02$ mag} & \multicolumn{2}{r}{1}\\
\multicolumn{2}{l}{[4.5]} & \multicolumn{2}{c}{$14.73\pm0.02$ mag} & \multicolumn{2}{r}{1}\\
\multicolumn{2}{l}{[5.8]} & \multicolumn{2}{c}{$14.39\pm0.05$ mag} & \multicolumn{2}{r}{1}\\
\multicolumn{2}{l}{[8.0]} & \multicolumn{2}{c}{$14.15\pm0.04$ mag} & \multicolumn{2}{r}{1}\\
\hline \hline
\multicolumn{6}{c}{Physical Properties}\\
\hline
\multicolumn{2}{l}{Age (Myr)} & \multicolumn{2}{c}{30--150}  & \multicolumn{2}{r}{1}  \\
\multicolumn{2}{l}{log($L_{\rm bol}/L_{\odot}$)} & \multicolumn{2}{c}{$-4.75\pm0.04$}   & \multicolumn{2}{r}{1}   \\
\multicolumn{2}{l}{Mass ($M_{\rm Jup}$)} & \multicolumn{2}{c}{11$^{+4}_{-3}$}   & \multicolumn{2}{r}{1}   \\
\multicolumn{2}{l}{$T_{\rm eff}$ (K)} & \multicolumn{2}{c}{$1040\pm50$}   & \multicolumn{2}{r}{1}   \\
\multicolumn{2}{l}{log($g$)} & \multicolumn{2}{c}{$4.2^{+0.2}_{-0.1}$}   & \multicolumn{2}{r}{1}   \\
\multicolumn{2}{l}{Near-IR spectral type} & \multicolumn{2}{c}{L7.5$\pm0.5$}   & \multicolumn{2}{r}{2}  \\
\multicolumn{2}{l}{Near-IR gravity class} & \multicolumn{2}{c}{{\sc VL-G}}   & \multicolumn{2}{r}{1}  
\enddata
\tablerefs{1. This paper; 2. \citet{2006ApJ...651.1166M}.}
\end{deluxetable}

\section{RESULTS AND ANALYSIS}\label{results}
\subsection{New constraints on the age of the system}

\citet{2006ApJ...651.1166M} determined that HD 203030 has a likely age of 250 Myr, with an age range of 130--400 Myr, which was also adopted as the likely age of HD 203030B. Several near-IR spectra of young brown dwarfs and planets have been collected during the last decade, and they suggest that HD 203030B could be younger than initially thought. In Section \ref{secundaria} we compare the new data collected for HD 203030B with those of other young ultra-cool dwarfs, and in Section \ref{primaria}, we re-assess the age of the primary using new self-consistent photometric data from various surveys over the past decade.

\subsubsection{Spectroscopic signatures of youth in HD 203030B  \label{secundaria}}

\begin{figure}
\centering
\includegraphics[width=0.49\textwidth]{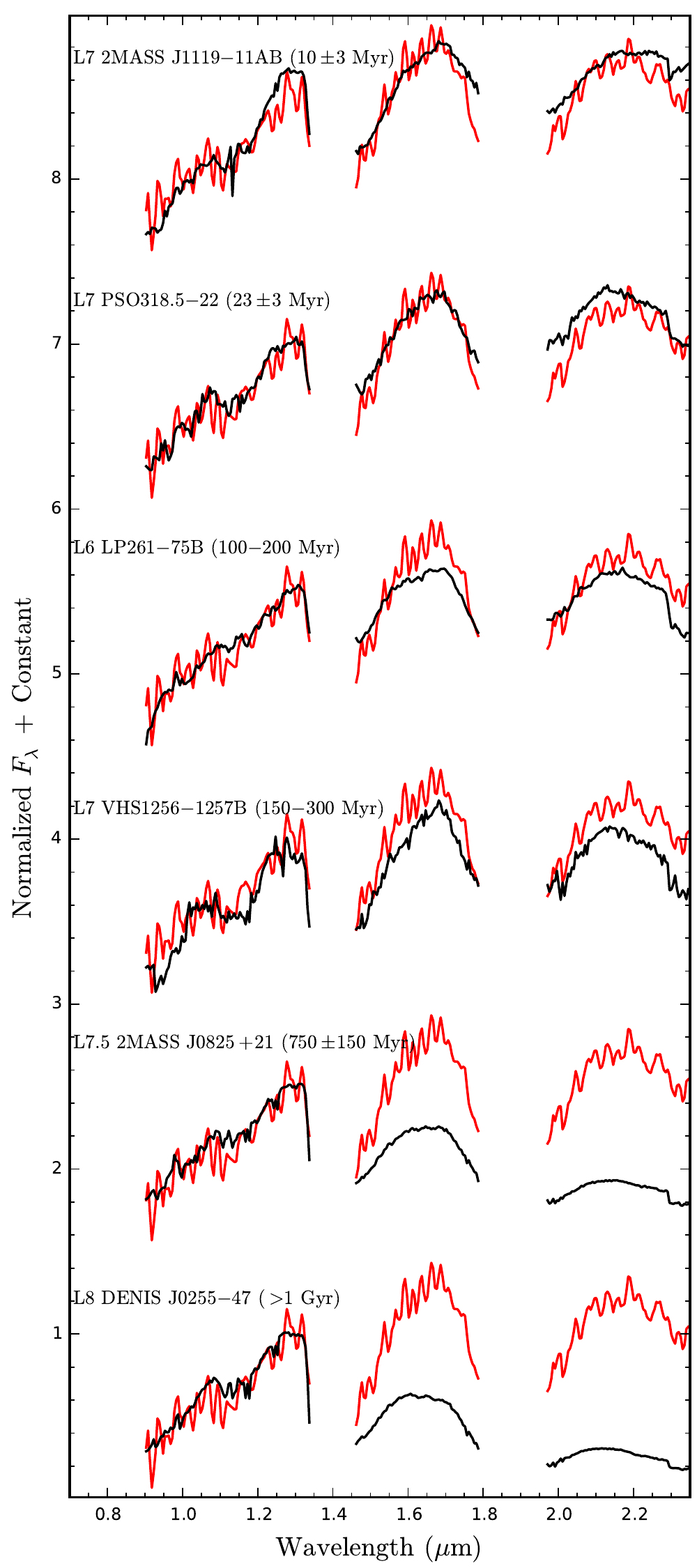}
\caption{ 
SpeX prism normalized spectra of HD 203030B (red) compared to spectra of other ultra-cool dwarfs (black) with similar spectral type. Comparison dwarfs are from top to bottom: 2MASS J11193254--1137466AB \citep{2016ApJ...821L..15K}, PSO J318.5338--22.8603 \citep{2013ApJ...777L..20L}, LP 261--75B \citep{2013ApJ...774...55B}, VHS J125601.92--125723.9 \citep{2015ApJ...804...96G}, and 2MASS J08251968$+$2115521 and DENIS-P J025503.3--470049 \citep{2005ApJ...623.1115C}. Spectra of comparison dwarfs have been degraded to the same spectral resolution as the spectra of HD 203030B.\label{figlow}}
\end{figure}

Independently of the age of HD 203030, we investigate the likely youth of HD 203030B by comparing its spectra to those of other ultra-cool dwarfs with a well determined age. In Figure \ref{figlow} we plot the low-resolution spectra of HD 203030B and other comparison ultra-cool dwarfs with different ages and similar spectral types. The $J$ band spectrum of all comparison objects and HD 203030B agree very well. The $H$ band spectrum of HD 203030B is redder and exhibits a more triangular shape than dwarfs with ages older than 600 Myr, and it is similar to the typical $H$ band of late-L dwarfs with ages in the range 10--300 Myr. A triangular-like $H$ band has been proposed as a sign of youth \citep{2001MNRAS.326..695L} if combined with other indicators \citep{2013ApJ...772...79A} since it can also be a sign of low metallicities \citep{2016AJ....151...46A}. In the $K$ band, HD 203030B is slightly redder than VHS J1256-12B \citep[150--300 Myr,][]{2015ApJ...804...96G}, and comparable to LP 261--75B \citep[100--200 Myr,][]{2006PASP..118..671R}, PSO J318.5--22 \citep[$23\pm3$ Myr,][]{2013ApJ...777L..20L,2016ApJ...819..133A}, or 2MASS J1119--11AB \citep[$10\pm3$ Myr,][]{2016ApJ...821L..15K,2017ApJ...843L...4B}. Thus, the low-resolution spectra of our target confirm the L7.5\,$\pm$\,0.5 spectral type assigned by \citet{2006ApJ...651.1166M}  and suggest an age in the 10--200 Myr range.
\

\begin{figure}
\centering
\includegraphics[width=0.49\textwidth]{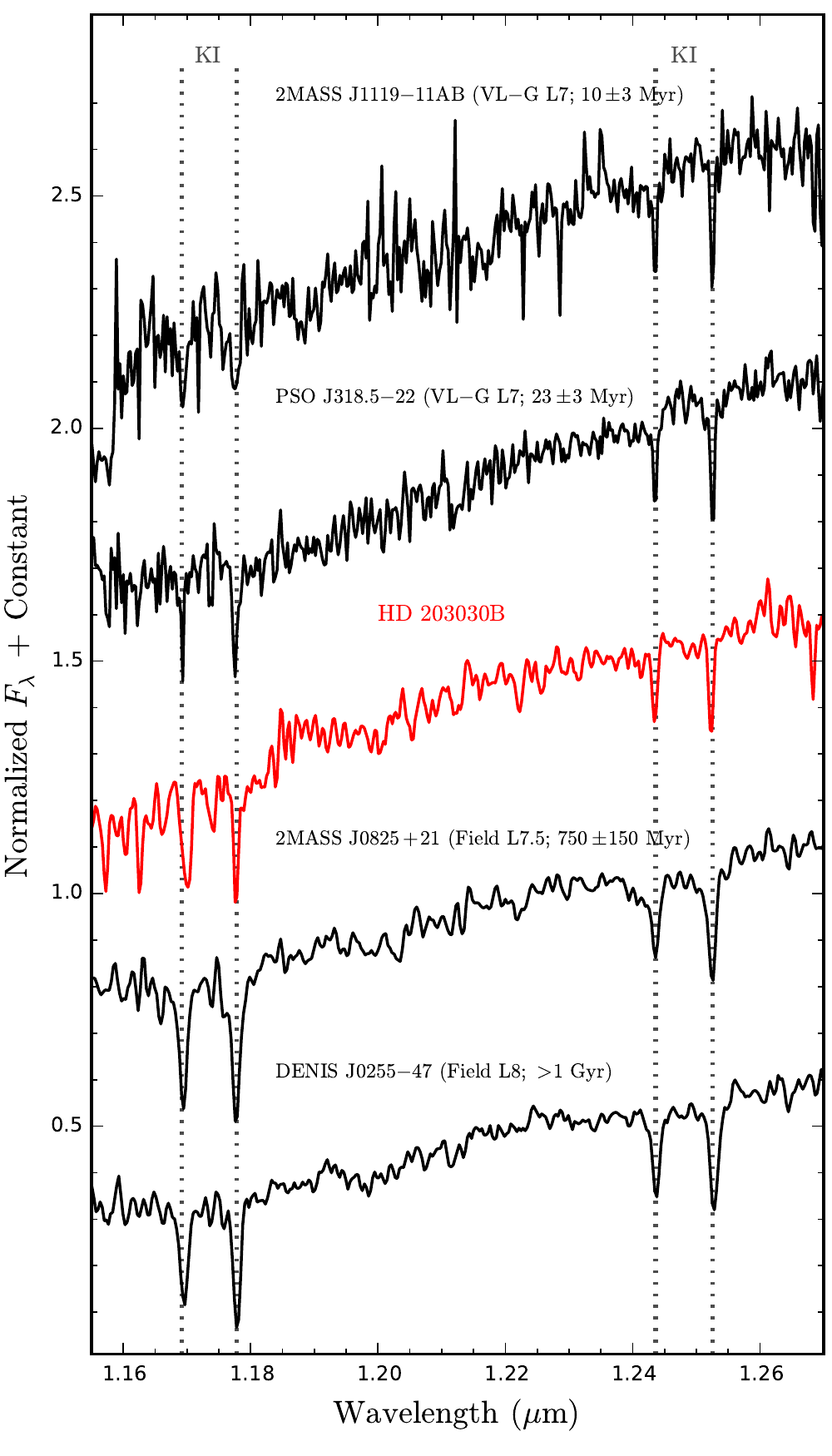}
\caption{$J$ band moderate-resolution spectra of HD 203030B (red) and some comparison dwarfs (black; degraded to the same spectral resolution as HD 203030B) from Figure \ref{figlow}. All spectra have been normalized and shifted in the $y$-axis for comparison purposes. \ion{K}{1} lines are indicated.
\label{figJband}}
\end{figure}

\begin{figure*}
\centering
\gridline{ \fig{fig_color_jjh.eps}{0.49\textwidth}{} \fig{fig_color_jjk.eps}{0.49\textwidth}{} }
\gridline{\fig{fig_color_jc2-spt.eps}{0.49\textwidth}{} \fig{fig_color_irac12-spt.eps}{0.49\textwidth}{}}
\caption{Color-Magnitude diagrams (top) and IR colors vs. spectral type using the compilation of ultra-cool dwarfs with known trigonometric parallaxes from \citet{2012ApJS..201...19D} and \citet{2016ApJ...833...96L}, and the recent findings of  \citet{2017ApJ...843L...4B} for 2MASS J11193254--113746AB. HD 203030B is shown in red. Ultra-cool dwarfs with low-gravity features in their spectra are plotted with green symbols. In the top panels, M, L and T dwarfs considered to be older than 500 Myr are plotted with black, gray and light gray symbols, respectively. For comparison purposes, we have included PSO J318--22 and 2MASS J11193254--113746AB in the bottom panels by transforming their $W1$ and $W2$ magnitudes to [3.6] and [4.5] following the relations given in \citet{2012ApJS..201...19D}. \label{figcolor}}
\end{figure*}

We also investigated the moderate-resolution spectra of HD 203030B, in particular, its $J$-band since it contains \ion{K}{1} lines that are sensitive to surface gravity. This spectrum and those of other L7-L8 dwarfs are plotted in Figure \ref{figJband}. Our $J$ band spectrum shows weak \ion{K}{1} lines at $\lambda$1.1778 $\mu$m, $\lambda$1.2437 $\mu$m, and $\lambda$1.2529 $\mu$m. We do not consider the \ion{K}{1} line at $\lambda$1.1692 $\mu$m since it appears unusually wide compared to the other three, which suggests that this line may be contaminated by noise: as seen at $\leq$1.165 $\mu$m. In addition, our spectrum also shows a weak FeH band at $\lambda$1.20 $\mu$m. From Figure \ref{figJband} it is evident that the \ion{K}{1} absorption lines of HD 203030B are comparable to those of objects in the range 10--100 Myr, and are narrower and weaker than those of dwarfs older than $\sim600$ Myr. We computed the equivalent widths (EW) of the \ion{K}{1} lines at $\lambda$1.177 (EW$ = 3.5\pm0.4$ \AA) and $\lambda$1.253 $\mu$m (EW$ = 1.9\pm0.3$ \AA), and the FeH$_J$ and $H$-cont indices following the prescription for gravity classification of \cite{2013ApJ...772...79A}, and classified HD 203030B as a very low-gravity object \citep[{\sc vl-g}; ages $\lesssim$150 Myr,][]{2013ApJ...772...79A,2016ApJ...833...96L}, which supports the age range derived from the low-resolution spectra.

In Figure \ref{figcolor} we plot the $J$ absolute magnitude as a function of the colors $J-H$ and $J-K$ (top panels) and the colors $J-[4.5]$ and $[3.6]-[4.5]$ vs. spectral type (bottom panels) for HD 203030B and other ultra-cool dwarfs taken from the literature. Our new measurement of the $J$ band magnitude of HD 203030B---about 7 times more accurate than the previous one---and the $H$ and $K$ band values yield near-IR colors $J-H\,=\,1.20\pm0.11$ mag and $J-K\,=\,2.56\pm0.13$ mag. The $J$, $H$ and $K$ band absolute magnitudes of HD 203030B are $15.77\pm0.08$, $14.58\pm0.08$, and $13.22\pm0.10$ mag, respectively, and are $\sim$1.1, $\sim$0.8, and $\sim$0.3 mag fainter than the measured for mature L7-L8 dwarfs. From the top panels of Figure \ref{figcolor}, HD 203030B is fainter and much redder than late-L field counterparts, and occupies a position similar to other young planetary-mass objects. A similar behavior is also seen in the bottom panels when comparing infrared colors versus spectral types.

\subsubsection{Re-assessment of the youth of HD 203030 \label{primaria}}

HD 203030 (G8V) was assigned an age of $\sim$250 Myr with a possible age range of 130--400 Myr \citep{2006ApJ...651.1166M}. This age estimation was obtained after inspection of the position of HD 203030 on a color-magnitude diagram---using $BV$ photometry from \citet{1964AJ.....69..570E} with no published uncertainties---in combination with other age indicators such as the stellar rotation, emission from chromospheric and coronal lines, lithium absorption, and the likely membership to young moving groups. We have revised and compared all of these age indicators with more recent observations and calibrations.

\begin{figure}[]
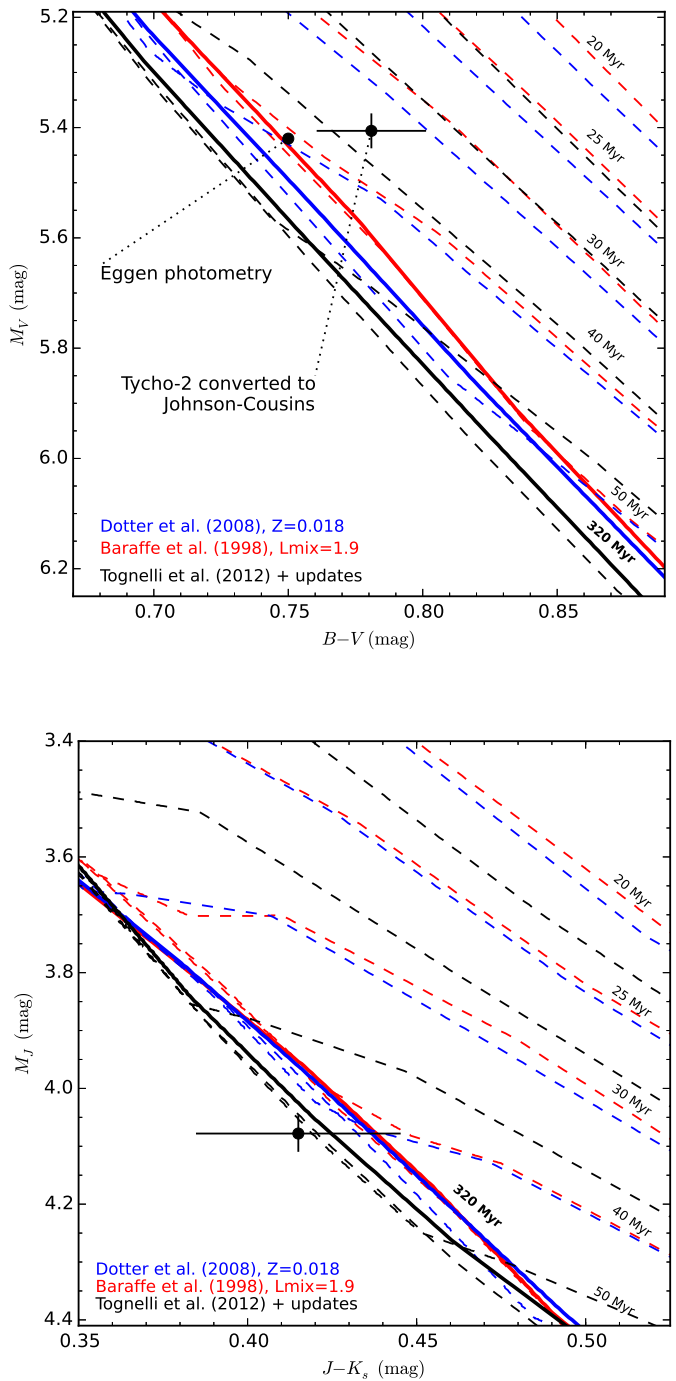

\centering
\gridline{\leftfig{fig_iso_opt.eps}{0.49\textwidth}{}}
\gridline{ \leftfig{fig_iso_nir.eps}{0.49\textwidth}{} }
\caption{Optical (top) and near-infrared (bottom) color-magnitude diagrams. The position of HD 203030 is indicated with a black data point. Vertical and horizontal bars indicate the uncertainties in the measurements. The Eggen optical photometry used in \citet{2006ApJ...651.1166M} and the Tycho-2 photometry used here are shown in the top panel. We have also included evolutionary tracks (red: \citealt{1998A&A...337..403B}; blue: \citealt{2008ApJS..178...89D}; black: \citealt{2011A&A...533A.109T,2012A&A...548A..41T}) for ages between 20 Myr and 50 Myr (dashed lines), and for 320 Myr (continuous lines).   \label{cmd}}
\end{figure}

We use Tycho-2 photometry \citep[$B_T=9.393\pm0.019$ mag, $V_T=8.513\pm0.013$ mag;][]{2000A&A...355L..27H} for HD 203030, and converted to the Johnson-Cousins system by using the relations from \citet{2002AJ....124.1670M}, for which there are well-established uncertainties. We obtain a Johnson $B-V$color of $0.781\pm0.024$ mag, and an apparent magnitude of $V=8.425\pm0.014$ mag, which in combination  with the Tycho-Gaia trigonometric parallax of HD 203030 \citep[$25.24\pm0.24$ mas;][]{2015A&A...574A.115M} leads to an absolute magnitude of $M_V = 5.436\pm0.030$ mag. The values of $M_V$ and $B-V$ are plotted in the top panel of Figure \ref{cmd}, for which we also plot the evolutionary tracks from \citet{1998A&A...337..403B}, \citet{2008ApJS..178...89D}, and \citet{2011A&A...533A.109T,2012A&A...548A..41T}. This diagram shows that HD 203030 occupies a position compatible with an age of 30--40 Myr, even though a main-sequence age is in agreement within 2$\sigma$. Similarly, using 2MASS photometry, $J=7.068\pm0.019$ mag and $K_s=6.653\pm0.023$ mag, we obtain a $J-K_s$ color of $0.415\pm0.030$ mag, and a $J$-band absolute magnitude of $M_J = 4.01\pm0.10$~mag. These values are plotted in the bottom panel of Figure \ref{cmd} with the same evolutionary tracks, and they show that HD 203030 is compatible with main sequence ages, but also with ages as young as 35~Myr. We note that the position in a color-magnitude diagram and the comparison with isochrones are not a unique indicator of age since there is a degeneracy between age and metallicity, as pointed out by \citet{2007ApJS..168..297T} and \citet{2013ApJ...776....4N}. Nonetheless, this result agrees with the likely younger age of HD 203030B.

Stellar rotation is another primary age indicator. \citet{2006ApJ...651.1166M} compared the rotation period of HD 203030, obtained by \citet[][$\sim$4.1 d]{2002MNRAS.331...45K} from Hipparcos photometry, to those observed for stars with similar spectral types in stellar associations with well known ages, and concluded that HD 203030 shows a rotation period compatible with those observed in $\alpha$ Per ($\sim$90 Myr) and the Hyades ($\sim$625 Myr). We note that there is a second rotation period measurement of $\sim$6.7 d reported by \citet[][based on spectroscopic observations]{2000A&AS..142..275S}, which is different from the one reported by \citet{2002MNRAS.331...45K}. Regardless of which value of the rotation period is correct, we used the rotation-color-age gyrochronology relation developed by \citet[][see their Figure 5]{2007ApJ...669.1167B} to infer the expected age of HD 203030 from its color $B-V$ and its rotation period. We find ages of $\sim$90 Myr and 240 Myr for the rotation period of 4.1 d and 6.7 d, respectively. \citet{2015A&A...584A..30L} also modelled the rotation periods observed for sun-like stars in stellar clusters with ages in the range $\sim$0.1--2.5 Gyr. From the results of these authors, we find an age of $\sim$100--160 Myr for rotation periods in the range $\sim$4--7 d and stellar masses in the range $\sim$0.95--1.00 M$_\odot$, as expected for HD 203030.

\begin{deluxetable}{lccccr}
\tabletypesize{\footnotesize}
\tablewidth{0pt}
\tablecaption{Summary of the ages of HD 203030A/B obtained from different estimators.
\label{tab_ages}}
\tablehead{\multicolumn{3}{l}{Method} & \multicolumn{3}{c}{Age (Myr)}}
\startdata 
\multicolumn{6}{c}{\bf HD203030B}\\
\multicolumn{2}{l}{near-IR low-resolution spectroscopy} &\multicolumn{2}{c}{ } & \multicolumn{2}{r}{10--200}\\
\multicolumn{2}{l}{near-IR moderate-resolution spectroscopy} &\multicolumn{2}{c}{ } & \multicolumn{2}{r}{$\le$150}\\
\multicolumn{6}{c}{\bf HD203030A}\\
\multicolumn{2}{l}{Isochrones} &\multicolumn{2}{c}{ } & \multicolumn{2}{r}{$\ge35$}\\
\multicolumn{2}{l}{Gyrochronology} &\multicolumn{2}{c}{ } & \multicolumn{2}{r}{90--240}\\
\multicolumn{2}{l}{$R^\prime_{HK}$ activity indicator} &\multicolumn{2}{c}{ } & \multicolumn{2}{r}{5--240}\\
\multicolumn{2}{l}{Kinematics} &\multicolumn{2}{c}{ } & \multicolumn{2}{r}{35--55}\\
\enddata
\end{deluxetable}

The new $R^\prime_{HK}$ value from \citet[][$R^\prime_{HK} = -4.370$]{2013A&A...552A..27M} is consistent with the one assembled in \citet{2006ApJ...651.1166M}.  We used the updated age-activity relationship from \citet{2008ApJ...687.1264M} and found a very consistent result: 240 Myr vs the $180^{+260}_{-140}$~Myr reported by \citet{2006ApJ...651.1166M}, even though \citet{2008ApJ...687.1264M} note that the scatter in the relationship at the ZAMS (Pleiades) corresponds to an order of magnitude uncertainty in age.  From Figure 4 of \citet{2008ApJ...687.1264M}, we find that the $R^\prime_{HK}$ value corresponding to HD 203030 is in the Pleiades domain, but it is also consistent with some of the less active members of the 5--25~Myr Sco-Cen OB association \citep[][and references therein]{2015MNRAS.448.2737R}.

Regarding kinematics, \citet{2001MNRAS.328...45M} proposed HD 203030 as a member of the IC 2391 supercluster \citep[35--55 Myr,][]{1981A&A....97..235M,2004ApJ...614..386B,2007A&A...461..509P,2008hsf2.book..757T,2009ApJ...706.1484B} based on the position of HD 203030 in the ($U$, $V$, $W$) space, and using the Eggen's kinematic criteria. We have computed the Galactic-space velocities $UVW$ of HD203030 following \citet{1987AJ.....93..864J} and by using the new values for its proper motion and trigonometric parallax from the Tycho-Gaia solution, and the new radial velocity derived by \citet{2013A&A...552A..64S}. We found ($-22.84\pm0.18$, $-15.52\pm0.02$, $-12.17\pm0.16$) km\,s$^{-1}$, which is in agreement with the values derived by  \citet{2001MNRAS.328...45M} and with the core velocity of IC 2391. However, the entity of the IC 2391 supercluster has been brought into question by \citet{2016IAUS..314...21M} based on the wide velocity dispersion of the members of this group, which suggests that these members were not near each other a few Myr ago. Therefore, we do not consider the kinematics a strong factor for confirming youth.  Nonetheless, we note that the $UVW$ motion of HD 203030 is consistent with the ``young field'' stellar population \citep{2013ApJ...762...88M,2014ApJ...783..121G}. This is still in good agreement with the results from Figure \ref{cmd}.

In conclusion, our use of more accurate photometry suggests ages younger than the 130--400 Myr adopted by \citet{2006ApJ...651.1166M}. Other age indicators, including activity, rotation, or kinematic membership to a young moving group continue to point to a wider range of ages, but are still compatible with ages younger than 100 Myr. A summary of the ages derived in Sections \ref{secundaria} and \ref{primaria} is presented in Table \ref{tab_ages}.

\subsection{A revision of the physical properties of HD 203030B}

From the age constraints of Sections  \ref{secundaria} and \ref{primaria} we adopt a revised age of 100 Myr with a likely range of 30--150 Myr for HD 203030 and HD 203030B. In this Section we revise the physical properties of the companion with this new age estimation.

\subsubsection{Luminosity}

We used our near-infrared spectra and the {\sl Spitzer} photometry to compute the luminosity of HD 203030B. To complete the spectral energy distribution (SED) of HD 203030B at wavelengths shorter than $\sim$0.85 $\mu$m and longer than $\sim$9 $\mu$m, we fit our data to different families of theoretical spectra such as Ames-DUSTY/COND \citep{2001ApJ...556..357A}, BT-COND/DUSTY/Settl \citep{2012RSPTA.370.2765A}, or Drift-PHOENIX \citep[][and references therein]{2008ApJ...675L.105H} by using the {\sl virtual observatory SED analyzer} \citep[VOSA,][]{2008A&A...492..277B}. We found that different combinations of $T_{\rm eff}$ in the range 1400--1600 K and surface gravities 3.5$\le$ log($g$) $\le$4.0 dex resulted in good fits for all explored models (assuming solar metallicity). We used these theoretical spectra to complete the SED of HD 203030B. Then, we integrated the SED in the range 0--1000 $\mu$m to obtain the bolometric luminosity from the equation $L_{\rm bol}\,=\, 4\pi\,d^{2}\,\int^{1000}_{0}F_{\lambda}d\lambda$, and found a value of log($L_{\rm bol}/L_{\odot}$)$\,=\,-4.75\pm0.04$ dex. We estimated the uncertainty on $L_{\rm bol}$ by propagating the errors on the parallax, our spectrophotometry, and the dispersion among the spectrophotometry of the theoretical spectra that best fit the 0.9--10 $\mu$m continuum. Separately, we integrated the SED of HD 203030B only in the range for which we have data---i.e., spectra in $\sim$0.85--2.4 $\mu$m plus the {\it Spitzer} photometry in the bands [3.6], [4.5], [5.8], and [8.0]---and found a pseudo-luminosity value of --4.76 dex, which is nearly 96 \% of the total bolometric luminosity derived by combining our data with theoretical spectra at wavelengths smaller than 0.85 $\mu$m and greater than $\sim$9 $\mu$m. The new bolometric luminosity is 1.4$\sigma$ lower than the value of \citet{2006ApJ...651.1166M}. The accuracy of the current value supersedes that of the previous determination, as that was based solely on the $K$-band photometry of HD 203030B and a bolometric correction.

\subsubsection{HD 203030B lies at the planetary mass boundary}

\begin{figure}
\centering
\includegraphics[width=0.49\textwidth]{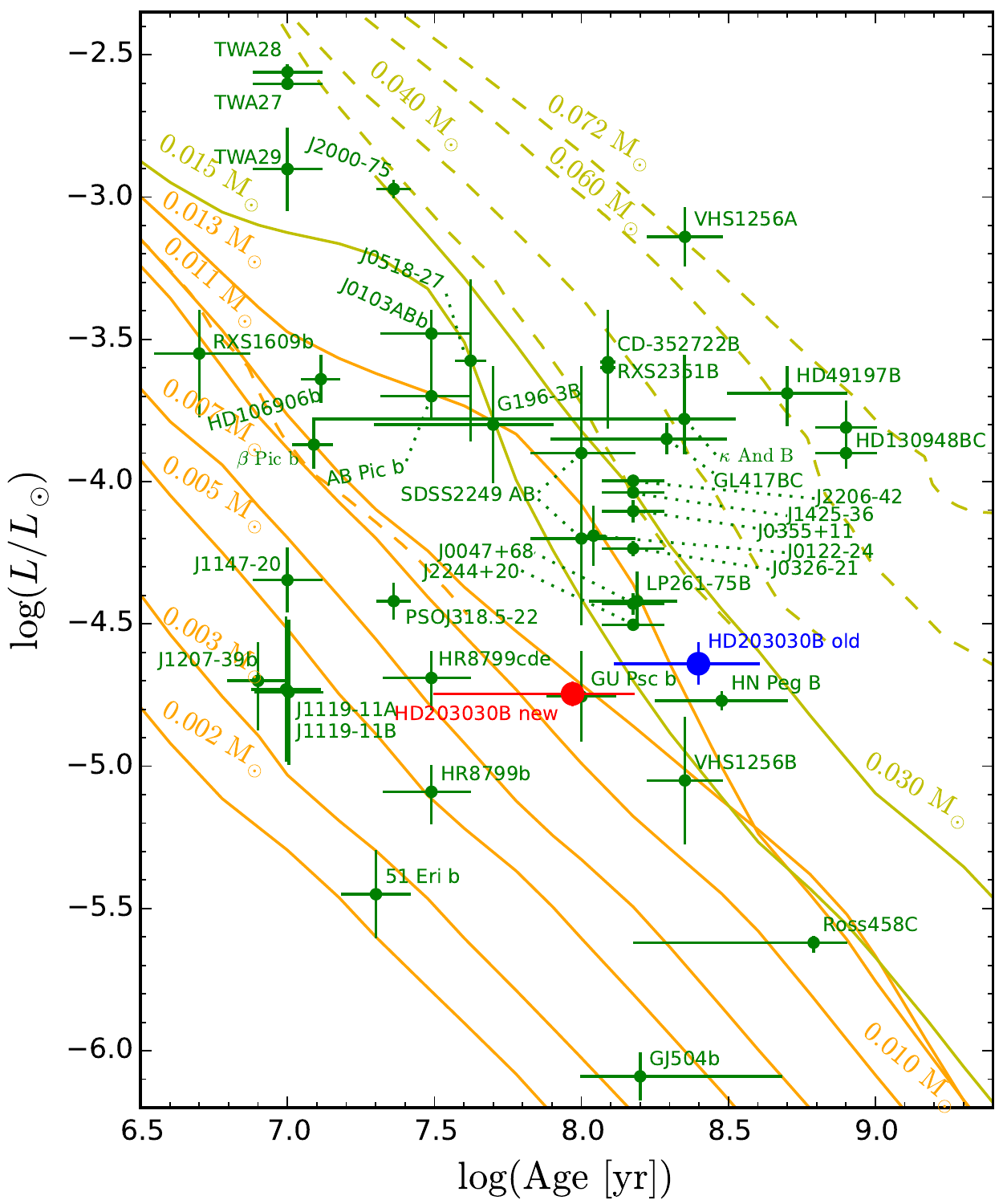}
\caption{Luminosity and age of HD 203030B derived in this work (red) and the previous determination from \citet[][blue]{2006ApJ...651.1166M}. Objects compiled by \citet{2013ApJ...774...55B}, \citet{2013ApJ...777L..20L}, \citet{2015ApJ...804...96G}, and \citet{2016ApJS..225...10F} are plotted with green symbols. Evolutionary tracks for dusty ultra-cool dwarfs from \citet[][continuous lines]{2008ApJ...689.1327S} and from \citet[][dashed lines]{2015A&A...577A..42B} are also plotted. Evolutionary tracks for objects less/greater than 0.013 M$_\odot$ are displayed in orange and yellow colors, respectively.
\label{figevo}}
\end{figure}

The new age estimate of HD 203030B (100 Myr with a range of 30--150 Myr) has a strong implication for its physical parameters, in particular its mass. In Figure \ref{figevo} we plot the bolometric luminosity as a function of age for several ultra-cool dwarfs, including HD 203030B with its previous age and luminosity estimates and with the ones obtained in this work. We have also included evolutionary tracks from \citet{2003A&A...402..701B} and \citet{2008ApJ...689.1327S}. The previous age estimate of HD 203030B showed it to be a 12--31 $M_{\rm Jup}$ brown dwarf as seen in Figure \ref{figevo}. Now, HD 203030B is immersed in the planetary mass regime with a mass of 11 $M_{\rm Jup}$ and a range of values of 8--15 $M_{\rm Jup}$ for the adopted age range.

Besides its mass, we have also explored the likely values of radius and surface gravity that evolutionary models predict for the observed luminosity and the new age range of HD 203030B. We find that for an age of 100 Myr, the typical radius predicted from \citet{2003A&A...402..701B,2015A&A...577A..42B} and \citet{2008ApJ...689.1327S} is 0.13 $R_{\odot}$ with a range of possible values of 0.126--0.136 $R_{\odot}$ for 30--150 Myr. We combined these model-dependent radii with our luminosity measurement to derive the $T_{\rm eff}$ of HD 203030B, and found a value of 1040$\pm50$ K, which is $\sim$400 K cooler than the $T_{\rm eff}$ of mature L7-L8 dwarfs \citep{2009ApJ...702..154S,2015ApJ...810..158F,2016ApJS..225...10F}. Our new $T_{\rm eff}$ is also lower than the reported value in the discovery paper of HD 203030B \citep[][$T_{\rm eff}=1206^{+74}_{-116}$ K]{2006ApJ...651.1166M}: a consequence of the  younger age and lower bolometric luminosity of the companion as determined here. Such unusually cool effective temperatures are now a known property of young planetary-mass objects \citep[e.g.,][ among others]{2007ApJ...654..570L,2013ApJ...777L..20L,2013ApJ...774...55B, 2015ApJ...804...96G}. Similarly for the surface gravity of HD 203030B, evolutionary models give a value of log($g$)$=4.2^{+0.2}_{-0.1}$ dex. The revised physical properties of HD 203030B obtained in Section \ref{results} are listed in Table \ref{tab_spitzer_phot}.

\section{CONCLUSIONS}\label{conclusions}

We obtained new infrared photometry and spectra of HD 203030B and compiled new photometric data of HD 203030 to better constrain their age and the physical properties of the companion. This revision was motivated by the new ability to independently assess the ages of young ultra-cool dwarfs, developed from the recent spectroscopic classification of substellar members of young moving groups \citep{2013ApJ...772...79A,2014A&A...562A.127B}.

Low-resolution spectra of HD 203030B are comparable to those of objects with ages in the 10--150 Myr range, which is also supported by the alkali lines and the $H$ band continuum of its moderate-resolution spectra. The near-IR absolute magnitudes of HD 203030B are fainter, and its $J-H$ and $H-K$ colors are redder, than those of $\gtrsim$1 Gyr field L7-L8 dwarfs.  On near-IR color-magnitude diagrams, HD 203030B sits close to other young L or T dwarfs thought to have planetary masses.  Independently, the optical and near-IR fluxes of the primary star HD 203030 indicate ages $\geq$30 Myr, while its activity and rotation point to an age in the $\sim$90--240 Myr range. HD 203030 is also a candidate member of the 35--50 Myr-old IC 2391 super cluster.  From the combined observational evidence, we adopt a new age of 100 Myr for HD 203030B, with 30--150 Myr as a possible range.

Our luminosity estimate combined with our new age estimation shows that HD 203030B has a mass of 11 $M_{\rm Jup}$, with a range of 8--15 $M_{\rm Jup}$, making it one of the first planetary-mass companions directly-imaged after 2MASS J12073346--3932539B \citep{2004A&A...425L..29C}. When combining the luminosity of HD 203030B with radii predicted from evolutionary models for the new age range, we find a $T_{\rm eff}$ of 1040$\pm50$ K, which is $\approx$400 K lower than the inferred for mature L7-L8 dwarfs. HD203030B is an excellent target to investigate the evolution of gaseous planetary-mass objects with the coming ground- and space-based telescopes. 

\acknowledgments

We are thankful to the anonymous referee for his/her very valuable comments. This work is based in part on observations made with the Spitzer Space Telescope, which is operated by the Jet Propulsion Laboratory, California Institute of Technology under a contract with NASA. Support for this work was provided by NASA  Jet Propulsion Laboratory under awards 1273192 
and 1369094.
Some of the data presented herein were obtained at the W.M.\ Keck Observatory, which is operated as a scientific partnership among the California Institute of Technology, the University of California and the National Aeronautics and Space Administration. The Observatory was made possible by the generous financial support of the W.M.\ Keck Foundation. The Center for Exoplanets and Habitable Worlds is supported by the Pennsylvania State University, the Eberly College of Science, and the Pennsylvania Space Grant Consortium. IRAF is distributed by the National Optical Astronomy Observatories, which are operated by the Association of Universities for Research in Astronomy, Inc., under cooperative agreement with the National Science Foundation. This work has also made use of data from the European Space Agency (ESA) mission {\it Gaia} (\url{https://www.cosmos.esa.int/gaia}), processed by the {\it Gaia} Data Processing and Analysis Consortium (DPAC, \url{https://www.cosmos.esa.int/web/gaia/dpac/consortium}). Funding for the DPAC has been provided by national institutions, in particular the institutions participating in the {\it Gaia} Multilateral Agreement.

\facility{Keck:II (NIRSPEC), IRTF (SpeX), Hale (WIRC), and Spitzer (IRAC).}
\software{IRAF \citep{1986SPIE..627..733T,1993ASPC...52..173T}, SpeXtool \citep[v3.2; ][]{2003PASP..115..389V,2004PASP..116..362C}, REDSPEC pipeline \citep{2003ApJ...596..561M}, IRAC Pipeline (V18.70), IRACProc \citep{2006SPIE.6270E..20S}, VOSA \citep{2008A&A...492..277B}.}

\bibliography{miles-paez_source}

\end{document}